\documentclass[pre,numbers,sort,twocolumn,showpacs,superscriptaddress,preprintnumbers,floatfix]{revtex4}

\usepackage{dcolumn}
\usepackage{url}
\usepackage{amsmath}
\usepackage{amssymb}
\usepackage{graphicx}
\usepackage{bm}   % bold math
\usepackage{bbm}
\usepackage{verbatim}
\usepackage{stmaryrd}
\usepackage{amsthm}
\newcommand\degrees[1]{\ensuremath{#1^\circ}}

\begin{document}

\title{Entomogenic Climate Change}

\author{David Dunn}
\email{artscilab@comcast.net}
\affiliation{Art and Science Laboratory, Santa Fe, New Mexico 87501}

\author{James P. Crutchfield}
\email{chaos@cse.ucdavis.edu}
\affiliation{Complexity Sciences Center \& Physics Department,
University of California Davis, One Shields Avenue, Davis, CA 95616}
\affiliation{Art and Science Laboratory, Santa Fe, New Mexico 87501}

\date{\today}

%\bibliographystyle{unsrt}

% ************************* ABSTRACT *************************
\begin{abstract}
Rapidly expanding insect populations, deforestation, and global climate change
threaten to destabilize key planetary carbon pools, especially the Earth's
forests which link the micro-ecology of insect infestation to climate. To the
extent mean temperature increases, insect populations accelerate deforestation.
This alters climate via the loss of active carbon sequestration by live trees
and increased carbon release from decomposing dead trees. A positive feedback
loop can emerge that is self-sustaining---no longer requiring independent
climate-change drivers. Current research regimes
and insect control strategies are insufficient at present to cope with the
present regional scale of insect-caused deforestation, let alone its likely
future global scale. Extensive field recordings demonstrate that bioacoustic
communication plays a role in infestation dynamics and is likely to be a critical
link in the feedback loop. These results open the way to novel detection and
monitoring strategies and nontoxic control interventions.
\end{abstract}


\preprint{Santa Fe Institute Working Paper 08-05-XXX}
\preprint{arxiv.org/q-bio.PE/0805XXX}

\maketitle

% ****************************************************************
\tableofcontents
%  ************************* INTRODUCTION *************************

\section{Introduction}

Forest ecosystems result from a dynamic balance of soil, plants, insects, animals, and
climate. The balance, though, can be destabilized by outbreaks of tree-eating insects.
These outbreaks in turn are sensitive to climate, which controls precipitation. Drought
stresses trees, rendering them vulnerable to insect predation. The net result is increased
deforestation driven by insects and modulated by climate.

For their part, many species of predating insects persist only to the extent that they
successfully reproduce by consuming and living within trees. Drought-stressed trees
are easier to infest compared to healthy trees, which have more robust defenses
against attack. To find trees suitable for reproduction, insects track relevant
environmental indicators, including chemical signals and, probably, bioacoustic
ones emitted by stressed trees. At the level of insect populations, infestation dynamics
are sensitive to climate via seasonal temperatures. Specifically, insect populations
increase markedly each year when winters are short and freezes less severe. The net
result is rapidly changing insect populations whose dynamics are modulated by climate.

Thus, via temperature and precipitation, climate sets the context for tree growth and
insect reproduction and also for the interaction between trees and insects. At the largest
scale, climate is driven by absorbed solar energy and controlled by relative fractions
of atmospheric gases. The amount of absorbed solar energy is determined by cloud and ground
cover. Forests are a prime example, as an important ground cover that absorbs, uses, and
re-radiates solar energy in various forms. At the same time forests are key moderators of
atmospheric gases. Trees exhaust oxygen and take up carbon dioxide in a process that
sequesters in solid form carbon from the atmosphere. As plants and trees evolved, in fact,
they altered the atmosphere sufficiently that earth's climate, once inhospitable, changed
and now supports a wide diversity of life.

\begin{figure}
\begin{center}
\resizebox{!}{2.50in}{\includegraphics{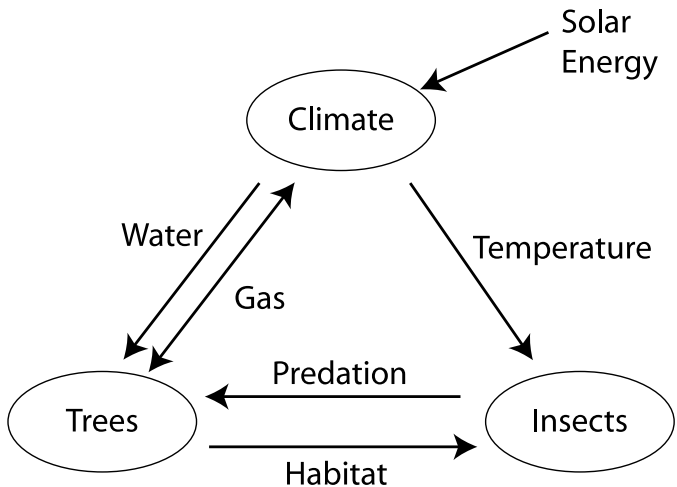}}
\end{center}
\caption{Insect, tree, and climate interactions discussed here; compare Net Primary
  Production \protect\citep{Fiel95a}.
  }
\label{fig:Interactions}
\end{figure}

There are at least three stories here: the trees', the insects', and the climate's. They
necessarily overlap since the phenomena and interactions they describe co-occur in space
and in time. Their overlap hints at an astoundingly complicated system, consisting of many
cooperating and competing components; the health of any one depending on the health of
others. (Figure \ref{fig:Interactions} gives a schematic view of the components and
interactions that we consider in the following; cf. \textcite{Fiel95a}.)
How are we to understand the individual views as part of a larger whole? In particular,
what can result from interactions between the different scales over which insects, trees,
and climate adapt?

Taking the stories together, we have, in engineering parlance, a \emph{feedback loop}: Going
from small to large scale, one sees that insects reproduce by feeding on trees; forests affect
regional solar energy uptake and atmospheric gas balance; and, finally, energy and carbon
storage and atmospheric gases affect climate. Simultaneously, the large scale (climate)
sets the context for dynamics on the small scale: temperature modulates insect
reproduction and precipitation controls tree growth. The feedback loop of insects, trees,
and climate means that new kinds of behavior can appear---dynamics not due to any
single player, but to their interactions. Importantly, such feedback loops can maintain
ecosystem stability or lead to instability that amplifies even small effects to large scale.

Here we give a concrete example of the dynamic interaction between insects, trees, and climate.
We focus on the role that bark beetles (Scolytidae or, more recently, Curculionidae: Scolytinae)
play in large-scale deforestation and consequently in climate change. Bark beetles are emblematic
of many different insect species that now participate in rapid deforestation. Likewise, we
primarily focus on the North American boreal forests for their unique characteristics but also
as representative of the vulnerability of all types of forest ecosystems. And so, the picture we
paint here is necessarily incomplete \footnote{An expanded discussion can be found at
\textcite{Dunn06c}.}.

Nonetheless, their cases serve to illustrate the complex
of interactions that are implicated in the feedback loop and also the current limits to human
response. Although they are not alone, bark beetles appear to be an example of a novel
player in climate change. Unlike the climatic role that inanimate greenhouse gases are
predicted to play in increasing global temperature over the next century, bark beetles
represent a biotic agent that actively adapts on the time scale of years but that, despite the
short time scale, still can cause effects, such as deforestation, at large spatial scales. To
emphasize the specificity and possible autonomy of this kind of biological, non-human agent,
we refer to the result as \emph{entomogenic climate change}.

A detailed analysis of the problem of entomogenic climate change leads us to make a
number of constructive suggestions for increased attention to relatively less familiar
domains of study, including micro-ecological symbiosis and its nonlinear population
dynamics, and insect social organization. Here we emphasize in particular the role that
bark beetle bioacoustic behavior must have in their evolving multiple survival
adaptations which, it appears, fill in significant gaps in the explanatory model of
infestation dynamics. One goal is to stimulate interdisciplinary research that is
appropriate to the complex of interactions implicated in deforestation and appropriate
to discovering effective control strategies.

\section{Forest Health and Climate}

The Earth's three great forest ecosystems---tropical, temperate, and boreal---are of
irreplaceable importance to its self-regulating balance. Their trees help to regulate its
climate, provide essential timber resources, and create a diversity of habitat and nutrients
that support other forms of life, including millions of people. Forests contribute to global
climate dynamics through a carbon cycle in which atmospheric carbon dioxide is converted
into an immense carbon pool. At any one point in time, the Earth's forest ecosystems together 
hold a majority of the terrestrial Earth's carbon stocks with the boreal forests comprising
49 percent of 
the total carbon pool contained within these three types of forest ecosystems \citep{Malh99a}. 
That carbon is then slowly released back into the atmosphere through complicated decomposition 
processes. 

All forms of deforestation, human and natural, directly impact climatic conditions by
attenuating or delaying the carbon cycle. In concert with well documented greenhouse
gas effects that drive global atmospheric change, the potential loss of large areas of
these forests, combined with accelerating deforestation of tropical and temperate regions,
may have significant future climate impacts beyond already dire predictions. Ice core
studies reveal that the Earth's climate has varied cyclically over the past 450,000 years.
Temperatures have been closely tied to variations in atmospheric carbon dioxide in a
cyclic change that recurs
on the time scale of millennia. Vegetation has been forced to adapt. The boreal forests are,
in fact, highly vulnerable to these climate shifts. Examination of fossil pollen and other
fossil records shows that, in response to temperature variations over the past millennia,
North American boreal forests have radically changed many times \citep{Lind02a}. The unique
sensitivity of these forests' tree species to temperature suggests that the predicted warmer
climate will cause their ecological niches to shift north faster than the forests can migrate.

One major consequence of boreal deforestation is increasing fire risk. Over the next
half-century, the Siberian and Canadian boreal forests will most likely see as much as a
50 percent increase in burnt trees \citep{Smit00a}. One of the major sources fueling these
fires will be dead and dying trees killed by various opportunistic insect species and their
associated microorganisms.

Paralleling concerns about the boreal forests, in recent years there has been a growing
awareness of extensive insect outbreaks in various regional forests throughout the
western United States. As consecutive summers of unprecedented forest fires consumed the dead
and dying trees a new concern emerged: insect-driven deforestation is a threat connected to
global climate change. In fact, climate experts, forestry personnel, and biologists, have all
observed that these outbreaks are an inevitable consequence of a climatic shift to warmer
temperatures \citep{Smit00a}.

Biologists now regularly voice concern that the problem exceeds any of the earlier
projections \citep{Kurz08a}. Evidence from diverse research sources suggests we are entering an
unprecedented planetary event: forest ecology is rapidly changing due to exploding
plant-consuming (\emph{phytophagous}) insect populations. In 2004, NASA's Global Disturbances
project analyzed nineteen years of satellite data ending in 2000. It revealed rapid
defoliation over a brief period (1995 to 2000) of a vast region that extends from the
US-Canadian border in western Canada to Alaska. The conclusion was that the
devastation resulted from two different insects, the mountain pine beetle (\emph{Dendroctonus
ponderosae}) and the western spruce budworm (\emph{Choristoneura occidentalis})
\citep{Pott05a}.

Now, four years later we know of even further damage. In Alaska, spruce bark beetles
(\emph{Dendroctonus rufipennis}) have killed 4.4 million acres of forest in just a
decade \citep{Berg06a}. This damage results from only one such insect. Climate warming
has also allowed the mountain pine beetle to expand its range into formerly unsuitable
habitats. The recent range expansion of the mountain pine beetle in British Columbia
has resulted in commercial timber losses of 435 million m$^3$, with additional losses
outside the commercial forests. The cumulative area of beetle outbreak was 130,000
km$^2$ by the end of 2006. This is an outbreak of unprecedented severity at a magnitude
larger in area than all previous recorded outbreaks \citep{Kurz08a}.

Jesse Logan (USDA Forest Service) and James Powell (Utah State University, Logan)
discussed the serious implications that a continuing warming trend will have on the
range expansion of the mountain pine beetle into both higher elevations and more
northern latitudes \citep{Loga01a}. At the time, one concern was that the beetles would breach the
Canadian Rockies and expand into the great boreal forests of Canada. Historically, these
forests have been immune to mountain pine beetles due to predictably severe winter conditions
that decimate beetle populations. Since much of Canada has seen mean winter
temperature increases as high as $\degrees{4}$C in the last century, and even faster changes
recently, the conditions for the beetles are improving rapidly.

It is now well established that mountain pine beetles have slipped through mountain
passes from the Peace River country in northern British Columbia to Alberta, the most
direct corridor to the boreal forests \citep{Carr06a}. If the beetle is successful at adapting
to and colonizing Canada's jack pine, there will be little to stop it moving through the immense
contiguous boreal forest, all the way to Labrador and the North American east coast. It
then will have a path down into the forests of eastern Texas. Entomologist Jesse Logan
\citep{Loga01a} describes this as ``a potential geographic event of continental scale with
unknown, but potentially devastating, ecological consequences.''

Continental migration aside, if the beetles infest the high-elevation conifers, the so-called
five-needle pines, of the western United States, this will reduce the snow-fence effect that
these alpine forests provide. Snow fences hold windrows of captured snow that are crucial to
the seasonal conservation and distribution of water from the Rocky Mountains. This is
one of the primary origins of water that sources several major river systems in North
America \citep{Loga01a}.
Every western state is contending with various rates of unprecedented insect infestation
not only by many different species of Scolytidae, but also by other plant-eating insects.

These and other rising populations of
phytophagous insects are now becoming recognized as a global problem and one of the
most obvious and fast emerging consequences of global climate change. Over the past
fifteen years there have been reports of unusual and unprecedented outbreaks occurring
on nearly every continent.

\section{What Drives Infestations?}

Several well-understood factors underlie how climate change impacts insect populations.
The two dominant environmental factors are changes in temperature and moisture.
Changing insect-host relationships and nonhost species impacts, such as predation and
disease, also play essential roles.

Since insects are cold-blooded, they are extremely sensitive to
temperature, being more active at higher temperatures. As winter temperatures increase,
there are fewer freezing conditions that keep insect populations in check than in the past.
Shortened winters, increasing summer temperatures, and fewer late-spring frosts correlate
to increased insect feeding, faster growth rates, and rapid reproduction \citep{Lomb00a}.

Moisture availability and variability are also major determinants of insect habitat---forest
health and boundaries. Drought creates many conditions that are favorable to increased
insect reproduction. Many drought-induced plant characteristics are attractive to insects.
Higher plant surface temperatures, leaf yellowing, increased infrared reflectance, biochemical
changes, and stress-induced cavitation acoustic emissions, may all be
positive signals to insects of host vulnerability. Drought also leads to increased food
value in plant tissues through nutrient concentration, while reducing defensive
compounds. These last factors may in turn increase the efficacy of insect immune
systems and therefore enhance their ability to detoxify remaining plant defenses. Higher
temperatures and decreased moisture may also decrease the activity of insect diseases and
predator activity while optimizing conditions for mutualistic microorganisms that benefit
insect growth \citep{Matt87a}.

One of the most frequently noted impacts of global climate change is the desynchronization
of biotic developmental patterns---such as the inability of forests to migrate as quickly
as their ecological niches---that have remained coherent for millennia. This decoupling
between various elements of an ecosystem is one of the most unpredictable and disruptive
results of abrupt climate change.

Unfortunately, insects respond to changes in their thermal environment much faster than
their hosts, either through migration (days), adaptation (seasonal), or evolution (centuries).
Under the stress of abrupt climate change the only short-term limit on their increasing
populations may be their near total elimination of suitable hosts. In short, trees only
adapt slowly (centuries) to changing conditions, while insects can disperse widely and
adapt much faster to abrupt environmental changes.

\section{The Tree's Perspective}

While it is clear that under extreme conditions phytophagous insects and their associated
microorganisms can quickly gain the advantage against host trees, it is also true that trees
have evolved effective defense mechanisms. For example, in their defense against bark
beetles there are two recognized components: the \emph{preformed resin system} and the \emph{induced
hypersensitivity response}. Once a beetle bores through the outer tree bark into the inner
tissues, resin ducts are severed and its flow begins. A beetle contends with the resin flow
by removing resin from its entrance hole. Trees that are sufficiently hydrated often
manage to ``pitch-out'' the invader through sufficient flow of resin. In some conifer
species with well defined resin-duct systems, resin is stored and available for beetle
defense. The \emph{monoterpenes} within the resin also have antibiotic and repellent properties
to defend against beetle-associated fungi \citep{Nebe93a}.

The induced hypersensitivity response is usually a secondary defense system; it is also
known as \emph{wound response}. It produces secondary resinosis, cellular dessication, tissue
necrosis, and wound formation---essentially a tree's attempt to isolate and deprive
nutrition to an invading organism. In species without well-defined resin-duct systems it is
often a primary defense mechanism. In both cases these defense strategies are very
susceptible to variations in temperature and available moisture. Their efficacy also varies
with different beetle species \citep{Nebe93a}.

Since winter survivability and the number of eggs laid by bark beetles is directly correlated
to ambient temperature \citep{Lomb00a}, it is no surprise that increases in
yearly beetle population cycles have been observed throughout the western states and
provinces as warming and local drought conditions have persisted \citep{Loga01a}.
The relative time scales for increased infestation rates, and subsequent adaptive tree
response, can put host trees at a serious disadvantage with regard to even the short-term
effects of climatic warming.

\section{Pioneer Beetle: Infestation Linchpin}

An attack begins with the pioneer beetle that locates, by means not yet elucidated, and lands
on a suitable host. Others join this beetle, all soon boring through the outer bark into
the phloem and cambium layers where eggs are laid after mating. Within the resulting
galleries that house the adult beetles and their eggs, the larvae hatch, pupate, and undergo
metamorphosis into adulthood. In this way, they spend the largest fraction of their life-cycle
(anywhere between 2 months to two years depending on species and geographic location) inside
a tree. This new generation emerges from the bark and flies away to seek new host trees.

The widely held view is that the pioneer attracts other beetles to the host through a
pheromone signal. Like many other insects, bark beetles manufacture communicative pheromones
from molecular constituents that they draw from host trees \citep{Agos92a}. In some species
the pioneer is male and, in others, female. Each new beetle that is attracted to the host
subsequently contributes to the general release of an \emph{aggregation} pheromone. It is
also theorized that the aggregation pheromone has an upper limit beyond which attracted beetles
will land on adjacent trees rather than the initial host, since high concentrations would indicate
over-use of the available host resources.

One hope has been that understanding bark beetle chemical ecology would lead to its
manipulation and eventually to a viable forestry management tool. Much to our loss,
nothing of the sort has been forthcoming. This largely derives from the sheer complexity
of the insect-tree micro-ecology and how far away we are from a sufficient understanding
of mechanisms and interactions. The two major contributions of chemical ecology
research to control measures have been those of pesticides and pheromone trapping. Most
biologists appreciate that pesticides have a very limited role in controlling insect
infestations at the scales in question. Pheromone traps are one of the essential tools of
field research in entomology, but adapting them for large-scale control has been
controversial at best; see \textcite{Bord97a} for an overview.

An underlying assumption of chemical ecology is that pheromones are the primary
attractant for beetles seeking new hosts, but this remains a hypothesis. While many
researchers believe that attraction is olfactory, others propose that visual cues are key for
some species \citep{Camp06a}. Importantly, forestry management policy is based largely on the
chemical ecology hypothesis that olfaction is dominant. It has never been definitively
proven, however, and, for a number of reasons, it is unlikely to be. Stated simply, foraging
insects most likely use whatever cues are the most accurate and easily assessed under varying
circumstances. To assume otherwise is to go against the common logic that living
systems evolve multiple survival strategies to cope with environmental complexity.

In short, key mechanisms in infestation dynamics remain unknown: the pioneer beetle's
ability to find a suitable host and then to facilitate organizing others to attack.

\section{The Bioacoustic Ecology Hypothesis}

One of the more under-appreciated research domains regarding bark beetles concerns their
remarkable bioacoustic abilities. The sound producing mechanism in many bark beetles is
a \emph{pars stridens} organ that functions as a friction-based grating surface.
In \emph{Ips confusus} beetles it is located on the back of the head and stroked by
a \emph{plectrum} on the under side of the dorsal anterior edge of the prothorax.
In other species (\emph{Dendroctonus}) the pars stridens is located on the surface
under the elytra and near the apices and sutural margins. Another is found in some
species on the underside of the head. All three of these sound generating organs
produce a variety of chirps that range from simple single-impulse clicks to a range
of different multi-impulse chirps. These also differ between genders of the same
species and between different species probably due to subtle differences in the
sound producing mechanisms. Collectively, all of the sounds and their associated
mechanisms are referred to as \emph{stridulation}, the most common form of sound
production made by various forms of beetle \citep{Barr69a}.

Past research suggested that sound making and perception in bark beetles was secondary
compared to their use of chemical-signaling mechanisms. Most studies addressing acoustic
behavior concentrated on sound generation and only in its relationship to chemical signaling.
These include the role stridulation sound-making has in controlling attack spacing between
entry points in the host \citep{Byer89a} or the triggering of pheromone release between genders
\citep{Rudi76a}. The resulting view is that bark beetles use a combination of chemical and
acoustic signals to regulate aggression, attack on host trees, courtship, mating behavior,
and population density.

While the dual behavioral mechanisms of scent and sound are largely inseparable, it is
usually assumed that bark beetles use chemical messages for communication at a distance
while reserving acoustic signals for close-range communication. However, this distinction
remains hypothetical. We do not yet have a
clear understanding of how far either their pheromones or sound signals can travel,
let alone a full appreciation of the diverse forms of acoustic signaling that they may
employ. We do know that both communication mechanisms are used after beetles have
aggregated on a host and that one form of signaling can evoke the other.

An emphasis on pheromone-based communication may very well have led to a lack of follow-up
on the possibility that host trees themselves produce acoustic cues that attract pioneer
beetles. Perhaps the earliest proposal dates to 1987, when William Mattson and Robert Haack
(USDA, Forest Service) speculated that cavitation events in trees might produce acoustic
signals audible to plant-eating insects \citep{Matt87a,Haac88a}. Cavitation occurs in trees by
breaking of the water columns in the conducting xylem tissue of leaves, stems, and trunks.
The assumption has been that the sounds are vibrations coming from individual cells
collapsing, which is due to gradual dehydration and prolonged water stress. While
cavitation produces some acoustic emissions in the audible range (20 Hz - 20 kHz), most occur
in the ultrasound range (20 - 200 kHz and above). In fact, counting ultrasonic acoustic emissions
from cavitating xylem
tissues is a widely accepted monitoring practice used by botanists to measure drought
stress in trees. Despite its common usage in botany, there has been very little study as to
the actual generating mechanism. For the most part, it is merely a statistical measuring
tool and the correlation between the incidence of cavitations and drought stress, an
accepted fact \citep{Jack96a}.

Recent fieldwork by one of us (DDD) focused on sound production by the pinion engraver
beetle (\emph{Ips confuses}). Sounds were recorded within the interior phloem layer of
the pinion trees, often adjacent to beetle nuptial chambers. A rich and varied acoustic ecology
was documented---an ecology that goes beyond the previously held assumptions about
the role of sound within this species \citep{Dunn06a}. Another important observation was
that much of the sound production by this species has a very strong ultrasonic component.
Since communication systems seldom evolve through investing substantial resources into a
portions of the frequency spectrum that an organism cannot both generate and perceive
\citep{Dunn06b}, this raised the question of whether or not bark beetles have a complementary
ultrasonic auditory capability. Recent laboratory investigations by Jayne Yack
(Biology, Carleton University) have also revealed ultrasound components in some bark beetle
signals and indirect evidence that beetles possess sensory organs for hearing airborne sounds
\citep{Yack06a}.

One possible implication that arises from the combination of these laboratory and field
observations is that various bark beetle species may possess organs capable of hearing
ultrasound for conspecific communication. If so, these species would be preadpated for
listening to diverse auditory cues from trees.

In turn, this raises an important issue not addressed by previous bark beetle
bioacoustic research. A very diverse range of sound signaling persists well after
the putatively associated behaviors---host selection, coordination of attack, courtship,
territorial competition, and nuptial chamber excavations---have all taken place. In
fully colonized trees the stridulations, chirps, and clicks can go on continuously for
days and weeks, long after most of these other behaviors will have apparently run their
course. These observations suggest that these insects have a more sophisticated social
organization than previously suspected---one that requires ongoing communication
through sound and substrate vibration.

The above acoustic fieldwork led us to conclude that there must be a larger range of
forms of insect sociality and so means of organizational communication. More precise
understanding of these forms of social organization may improve our ability to design
better control systems, whether these are chemical, acoustic, or biological.

The results in both bioacoustics and chemical ecology strongly suggest bark beetle
communication is largely multimodal and that both pheromone and mechanical signaling
are interwoven. A growing appreciation in many fields of biology has emerged that
animal signals often consist of multiple parts within or across sensory modalities. Insects
are not only an example of this observation, but they possess some of the most surprising
examples of multicomponent and multimodal communication systems \citep{Skal05a}.

\section{Conclusion: Closing the Loop}

The eventual impact that insect-driven deforestation and global climate change will have
on the Earth's remaining forests ultimately depends on the rate at which temperatures
increase. The impacts will be minimized if that rate is gradual, but increasingly disruptive
if the change is abrupt. Unfortunately, most climate projections now show that a rapid
temperature increase is more likely \citep{Wats01a}. The current signs of increasing insect
populations at this early stage of warming does not portend well for forest health in the
near future. The concern is exacerbated, since we have limited countermeasures under
development.

One conclusion appears certain. Extensive deforestation by insects will convert the essential
carbon pool provided by the Earth's forests into atmospheric carbon dioxide. Concomitantly,
the generation of atmospheric oxygen and sequestration of carbon by trees will decrease
\citep{Kurz08a}.

Most immediately, though, as millions of trees die, they not only cease to participate in the
global carbon cycle, but become potential fuel for more frequent and increasingly
large-scale fire outbreaks. These fires will release further carbon dioxide into the
atmosphere and do so more rapidly than the natural cycle of biomass decay. The interactions
between these various components and their net effect are complicated at best---a theme
running throughout the entire feedback loop.

An example of this is how boreal forest fires affect climate \citep{Rand06a}. A
constellation of substantially changed components (lost forest, sudden release of gases,
and the like) leads, it is claimed, to no net climate impact. The repeated lesson of
complex, nonlinear dynamical systems, though, is that the apparent stability of any part
can be destabilized by its place in a larger system. Thus, one needs to evaluate the lack of
boreal fire-climate effects in the context of the entire feedback loop.

Taken alone, the potential loss of forests is of substantial concern to humans. When
viewing this system as a feedback loop, though, the concern is that the individual
components will become part of an accelerating positive feedback loop of sudden
climatic change. Such entomogenic change, given the adaptive population dynamics of a
key player (insects), may happen on a very short time scale. This necessitates a shift in
the current characterization of increasing insect populations as merely symptomatic of
global climate change to a concern for insects as a significant generative agent.

In addition to concerted research in bioacoustics, micro-ecological symbiosis and
dynamics, and insect social organizations, these areas, in conjunction with the field of
chemical ecology, must be integrated into a broader view of multiscale population,
evolutionary, and climate dynamics. In this sense, the birth of chemical ecology serves as
an inspiration. It grew out of an interdisciplinary collaboration between biology and
chemistry. It is precisely this kind of intentional cooperation between disciplines---but
now over a greater range of scales---that will most likely lead to new strategies for
monitoring and defense against what seems to be a growing threat to the world's forests
and ultimately to humanity itself.

\section*{Acknowledgments}

The authors thank Dawn Sumner, Jim Tolisano, Richard Hofstetter, Jayne Yack,
Reagan McGuire, and Bob Harrill for helpful discussions. This work was partially
supported by the Art and Science Laboratory via a grant from the Delle Foundation
and the Network Dynamics Program, funded by Intel Corporation, at UCD and the
Santa Fe Institute.

\bibliography{ref}

\section*{Biography: James P. Crutchfield}

Jim Crutchfield teaches nonlinear physics at the University of California, Davis, directs
its Complexity Sciences Center, and promotes science interventions in nonscientific settings.
He's mostly concerned with what patterns are, how they are created, and how intelligent agents
discover them; see \url{cse.ucdavis.edu/~chaos}.

\section*{Biography: David D. Dunn}

David Dunn is a composer who rarely presents concerts or installations, preferring to lecture and
engage in site-specific interactions or research-oriented activities. Much of his current work
focuses on the development of strategies and technologies for environmental sound monitoring in
both aesthetic and scientific contexts.

\end{document}